\pdfoutput=1

\documentclass[aps,preprintnumbers,amsmath,amssymb,twocolumn, tightenlines,superscriptaddress]{revtex4}

\usepackage{graphicx}
\usepackage{epsfig}
\usepackage{float}
\usepackage{color}

\newcommand{\be}{\begin{eqnarray}}
\newcommand{\ee}{\end{eqnarray}}

 \newcommand{\gsim}{\mathrel{\hbox{\rlap{\lower.55ex \hbox {$\sim$}}
                   \kern-.3em \raise.4ex \hbox{$>$}}}}
\newcommand{\lsim}{\mathrel{\hbox{\rlap{\lower.55ex \hbox {$\sim$}}
                   \kern-.3em \raise.4ex \hbox{$<$}}}}

\newcommand{\ba}{\begin{eqnarray}}
\newcommand{\ea}{\end{eqnarray}}

\begin{document}


\title{Out-of-Equilibrium Chiral Magnetic Effect from Chiral Kinetic Theory}
\author{Anping Huang} 
\address{Physics Department, Tsinghua University, Beijing 100084, China.}
\address{Physics Department and Center for Exploration of Energy and Matter,
Indiana University, 2401 N Milo B. Sampson Lane, Bloomington, IN 47408, USA.}
\author{Yin Jiang} 
\address{Institut fuer Theoretische Physik, Philosophenweg 16, D-69120 Heidelberg, Germany.}
\author{Shuzhe Shi} 
\address{Physics Department and Center for Exploration of Energy and Matter,
Indiana University, 2401 N Milo B. Sampson Lane, Bloomington, IN 47408, USA.}
\author {Jinfeng Liao} \email{liaoji@indiana.edu}
\address{Physics Department and Center for Exploration of Energy and Matter,
Indiana University, 2401 N Milo B. Sampson Lane, Bloomington, IN 47408, USA.}
\author{Pengfei Zhuang} \email{zhuangpf@mail.tsinghua.edu.cn}
\address{Physics Department, Tsinghua University, Beijing 100084, China.}
\date{\today}

\begin{abstract}
Recently there has been significant interest in the macroscopic manifestation of chiral anomaly in many-body systems of chiral fermions. A notable example is the Chiral Magnetic Effect (CME). Enthusiastic efforts have been made to search for the CME in the quark-gluon plasma created in heavy ion collisions. A crucial challenge is that the extremely strong magnetic field in such collisions may last only for a brief moment and the CME current may have to occur at so early a stage that the quark-gluon matter is still far from thermal equilibrium. This thus requires  modeling of the CME in an out-of-equilibrium setting. With the recently developed  theoretical tool of chiral kinetic theory, we make a first phenomenological study of the CME-induced charge separation during the pre-thermal stage in heavy ion collisions. The effect is found to be very sensitive to the time dependence of the magnetic field and also influenced by the initial quark momentum spectrum as well as the relaxation time of the system evolution toward thermal equilibrium. Within the present approach, such pre-thermal charge separation is found to be modest.
\end{abstract}

\maketitle

\section{Introduction}

Spin-$\frac{1}{2}$ fermions that are massless (or approximately so)  are unique in that the axial symmetry in their classical description gets broken in quantized theory, a fundamental feature known as the chiral anomaly~\cite{Adler:1969gk,Bell:1969ts}. Examples of such chiral fermions include e.g. light quarks/leptons in the Standard Model of Particle Physics or emergent quantum states of electrons in the so-called Dirac and Weyl semimetals.  Recently there has been a rapidly growing interest in understanding the implications of microscopic quantum anomaly on the macroscopic properties of chiral matter, i.e. many-body systems with chiral fermions. It has been found that the chiral anomaly leads to a number of anomalous chiral transport processes that are absent in normal materials, such as the Chiral Magnetic Effect~\cite{Kharzeev:2004ey,Kharzeev:2007tn,Kharzeev:2007jp,Fukushima:2008xe}, Chiral Magnetic Wave~\cite{Kharzeev:2010gd,Burnier:2011bf}, Chiral Vortical Effects~\cite{Son:2009tf,Kharzeev:2010gr,Sadofyev:2010is,Landsteiner:2011iq,Jiang:2015cva}, etc. Enthusiastic efforts have been made to search for manifestation of these effects in two known types of chiral matter, the quark-gluon plasma (QGP) produced in heavy ion collisions~\cite{STAR_LPV1,STAR_LPV_BES,ALICE_LPV,Adamczyk:2015eqo} as well as the Dirac and Weyl semimetals~\cite{Li:2014bha,Xiong:2015nna,2015PhRvX...5c1023H,2016NatCo...711615A}. See recent reviews in e.g.~\cite{Kharzeev:2015znc,Kharzeev:2015kna,Liao:2014ava,burkov}

Let us focus on the  Chiral Magnetic Effect (CME), which predicts the generation of an electric current    $\vec{\bf J}_Q$ along the  magnetic field $\vec{\bf B}$ applied to the system, i.e.
\begin{eqnarray} \label{eq_cme}
\vec{\bf J}_Q = \sigma_5 \vec{\bf B}
\end{eqnarray}
where $\sigma_5 $ is the chiral magnetic conductivity, taking a universal value of $\frac{\mu_5 Q_f^2}{4\pi^2}$ in thermal equilibrium (for each species of chiral fermions), with $\mu_5$ the chiral chemical potential  that quantifies the imbalance between fermions of opposite (right-handed, RH versus left-handed, LH) chirality.  The search of CME in QGP has proven difficult, due to the complicated environment in heavy ion collisions. One major issue is that the magnetic field $\vec{B}$, necessary for driving the CME current (\ref{eq_cme}), is provided by the fast moving ions themselves. Such magnetic field has been extensively studied~\cite{Bzdak:2011yy,Deng:2012pc,Bloczynski:2012en,McLerran:2013hla,Gursoy:2014aka,Tuchin:2015oka,Li:2016tel,Inghirami:2016iru,Guo:2015nsa} and found to likely last only for a short time, i.e. on the order of $\lsim 1\rm fm/c$, after the impact of the two colliding ions. There is the possibility that the electrically conducting partonic matter created in the collision would develop induction current (upon deceasing magnetic field) and thus considerably elongate its duration. Whether this may quantitatively work, remains to be seen.

This situation therefore poses a challenge: during the time when the magnetic field is substantial (thus the CME current most significant), the partonic matter could still be far from thermal equilibrium. The quantitative modeling of CME in heavy ion collisions has only been recently achieved for the hydrodynamic stage and after~\cite{Jiang:2016wve,Yin:2015fca,Hirono:2014oda,Yee:2013cya}, so there remains a crucial gap to fill, namely quantifying the CME during the out-of-equilibrium stage in heavy ion collisions. In this paper, we aim to make an important step forward in addressing this pressing problem for the search of CME in QGP. Equipped with the  theoretical tool of chiral kinetic theory, we make a first phenomenological study of the CME-induced charge separation in the pre-thermal stage and discuss its implication for subsequent hydrodynamic evolution in heavy ion collisions.

\section{The Chiral Kinetic Theory (CKT)}

The theoretical framework to describe anomalous chiral transport in an out-of-equilibrium system is the recently developed chiral kinetic theory~\cite{Stephanov:2012ki,Son:2012wh,Son:2012zy,Chen:2014cla,Kharzeev:2016sut,Chen:2012ca,Hidaka:2016yjf,Mueller:2017lzw,Gorbar:2017cwv}. By introducing $\hat{O}(\hbar)$ correction to the classical equations of motion, it incorporates chiral anomaly related effects into the usual kinetic equation. In the following we present the necessary ingredients of this theory to be used for the present study. The kinetic equation to be solved takes the following form~\cite{Son:2012wh,Son:2012zy}:
 \begin{eqnarray}
 \left[ \partial_t + {\dot{\vec{\bf x}}}\cdot   {\vec{\nabla}}_{x}  +  {\dot{\vec{\bf p}}} \cdot \vec{\nabla}_{p} \right] f_i(\vec{\bf x},\vec{\bf p},t) = C[f_i]  \,\, ,
 \end{eqnarray}
 \begin{eqnarray} \label{eq_eom}
{\dot{\vec{\bf x}}} =  \frac{1}{\sqrt{G}} \left[{\vec{\bf v}_p}+ q_i\, \vec{\bf B} \, \left(  {{\vec{\bf v}_p}}\cdot \vec{\bf b} \right) \right] \,\, , \,\,   {\dot{\vec{\bf p}}} =  \frac{1}{\sqrt{G}} \left[ q_i\, {\vec{\bf v}_p}\times \vec{\bf B} \right] \, .\,
\end{eqnarray}
In the above, the $f_i$ is the distribution function in the phase space of position $\vec{\bf x}$ and momentum $\vec{\bf p}$, for each specie (labelled by $i$) of chiral fermions. The $C[f]$ is the collision term. The important change as compared with classical kinetic theory, is in the equations of motion (\ref{eq_eom}). 
Note here we consider the case with spatially homogeneous external magnetic field $\vec{\bf B}$ but without electric field, and $q_i$ is the electric charge of the type-$i$ fermions. 
In the above, an anomalous velocity along magnetic field $ q_i\, \vec{\bf B} \, \left(  {{\vec{\bf v}_p}}\cdot \vec{\bf b} \right) $ appears, with the Berry curvature $\vec{\bf b}\equiv \chi\, {\hat{\bf p}}/(2p^2)$ (where $p=|\vec{\bf p}|$ and $\chi=+/-$ for positive/negative helicity respectively).  The factor $G\equiv (1+ q_i\,  \vec{\bf B} \cdot \vec{\bf b})^2$ is a phase space factor. Another important effect of the $\vec{\bf{B}}$ field is the magnetization, which modifies particle dispersion into $\epsilon(\vec{\bf p})=p [1-\frac{g}{2} q_i \vec{\bf B}\cdot \vec{\bf b}]$ with $g$ the g-factor for magnetic moment for which we adopt the standard choice of $g=2$~\cite{Son:2012wh,Son:2012zy,Chen:2014cla}. The corresponding velocity $\vec{\bf v}_p=\vec{\bigtriangledown}_p\, \epsilon(\vec{\bf p})$ is thus given by 
\begin{eqnarray}
&&\vec{\bf{v}}_{p}=\hat{\bf{p}}\left(1+g q_i\vec{\bf B}\cdot\vec{\bf b}\right)-\frac{g}{2}q_i\vec{\bf B}\left(\hat{\bf p}\cdot\vec{\bf b}\right).
\end{eqnarray}
Substituting the above into the equation of motion and keeping up to linear order terms in $\vec{B}$, one obtains~\cite{Kharzeev:2016sut}
\begin{eqnarray} \label{eq_eom2}
 && \sqrt{G} \, {\dot{\vec{\bf x}}} =   \hat{\bf p}\left(1+gq_i\vec{\bf B}\cdot\vec{\bf b}\right) + a\, q_i\, \vec{\bf B} \, \left(  \hat{\bf p}\cdot \vec{\bf b} \right)  \,\, , \\
 &&  \sqrt{G} \, {\dot{\vec{\bf p}}} = q_i\, {\hat{\bf p}}\times \vec{\bf B}  \,\, \label{eq_eom3}
\end{eqnarray}
where $a=(2-g)/2$ vanishes for $g=2$. Finally the local current density can be obtained via the following:
\begin{eqnarray} \label{eq_ckt_j}
\vec{\bf j}_i = \int_{\vec{\bf p}}  \sqrt{G} \, {\dot{\vec{\bf x}}} \, f_i = \int_{\vec{\bf p}} \left[ {\vec{\bf v}_p}+ q_i\, \vec{\bf B} \, \left(  {{\vec{\bf v}_p}}\cdot \vec{\bf b} \right)\right]\, f_i  
\end{eqnarray}
with $\int_{\vec{\bf p}}\equiv \frac{d^3\vec{\bf p}}{(2\pi)^3}$. If one inserts a thermal Fermi-Dirac distribution in the above, the equilibrium CME current (\ref{eq_cme}) is reproduced. For details of CKT, see \cite{Stephanov:2012ki,Son:2012wh,Son:2012zy,Chen:2014cla}.

\section{Solutions to the CKT Equation}

Given the above chiral kinetic theory, it is of great interest to find possible analytic solutions. In the following we consider solutions for two different cases of the collision term, with a magnetic field $\vec{\bf B}=B(t) \hat{\bf y}$  along $y$-axis that has  constant magnitude across space but is time-dependent.

The first case is the collision-less limit, i.e. $C[f_i]=0$. In this case the particles will simply undergo ``free''-streaming according to the trajectory determined from the equations of motion (\ref{eq_eom2})(\ref{eq_eom3}). Note that such trajectory is different from the usual classical trajectory due to the anomalous terms. For a particle with initial position $\vec{\bf x}_0$ and initial momentum $\vec{\bf p}_0$ at time $t_0$, its position and momentum at a later time $t$ are given by: 
\begin{eqnarray}\begin{split}
&z=z_0+\frac{p_{z0}}{p}\int^{t}_{t_0}\frac{\zeta}{\sqrt{G}}\cos\theta' dt'+\frac{p_{x0}}{p}\int^{t}_{t_0}\frac{\zeta}{\sqrt{G}}\sin\theta' dt',\\
&x=x_0-\frac{p_{z0}}{p}\int^{t}_{t_0}\frac{\zeta}{\sqrt{G}}\sin\theta' dt'+\frac{p_{x0}}{p}\int^{t}_{t_0}\frac{\zeta}{\sqrt{G}}\cos\theta' dt', \,\, \\
&y=y_0+\frac{p_{y0}}{p}\int^{t}_{t_0}\frac{\zeta}{\sqrt{G}} dt'+\frac{a\, \chi}{2p}\int^{t}_{t_0}\frac{q_iB(t)}{p\sqrt{G}} dt'.
\end{split}\end{eqnarray}
\begin{eqnarray}\begin{split}
&p_{z}=p_{z0}\cos\theta+p_{x0}\sin\theta,\\
&p_{x}=-p_{z0}\sin\theta+p_{x0}\cos\theta,\\
&p_{y}=p_{y0}.
\end{split}\end{eqnarray}
In the above $\zeta=1+g q_i\vec{\bf B}\cdot\vec{\bf b}=1+g\chi\frac{q_iB(t)p_{y}}{2p^{3}}$, $\theta'=\int^{t'}_{t_0} \frac{\zeta\, q_iB(t')}{p\sqrt{G}} dt'  $,  $\sqrt{G}=1+q_i\vec{\bf{b}}\cdot\vec{\bf B}=1+\chi\frac{q_i\vec{p}\cdot\vec{\bf B}}{2p^{3}}=1+\chi\frac{q_iB(t)p_{y}}{2p^{3}}$.

Equivalently, a particle found to have position $\vec{\bf x}$ and momentum $\vec{\bf p}$ at a time $t$ can be traced back to a state of $\vec{\bf x}_0(\vec{\bf x},\vec{\bf p},t)=\vec{\bf x}(\vec{\bf x},\vec{\bf p},t; t0)=(x_0,y_0,z_0)$ and $\vec{\bf p}_0(\vec{\bf x},\vec{\bf p},t)=\vec{\bf p}(\vec{\bf x},\vec{\bf p},t; t0)=(p_{x0}, p_{y0}, p_{z0})$ at initial time, given by:
\begin{eqnarray}\begin{split}\label{eq:001}
&z_0=z-\frac{p_{z0}}{p}\int^{t}_{t_0}\frac{\zeta}{\sqrt{G}}\cos\theta' dt'-\frac{p_{x0}}{p}\int^{t}_{t_0}\frac{\zeta}{\sqrt{G}}\sin\theta' dt',\\
&x_0=x+\frac{p_{z0}}{p}\int^{t}_{t_0}\frac{\zeta}{\sqrt{G}}\sin\theta' dt'-\frac{p_{x0}}{p}\int^{t}_{t_0}\frac{\zeta}{\sqrt{G}}\cos\theta' dt', \,\,\,\, \\
&y_0=y-\frac{p_{y0}}{p}\int^{t}_{t_0}\frac{\zeta}{\sqrt{G}} dt'-\frac{a\, \chi}{2p}\int^{t}_{t_0}\frac{q_iB(t)}{p\sqrt{G}} dt'.
\end{split}\end{eqnarray}
\begin{eqnarray}\begin{split}\label{eq:002}
&p_{z0}=p_{z}\cos\theta-p_{x}\sin\theta,\\
&p_{x0}=p_{z}\sin\theta+p_{x}\cos\theta, \\
&p_{y0}=p_{y}.
\end{split}\end{eqnarray}

Therefore given an initial condition $f_{i\, 0}(\vec{\bf x}_0,\vec{\bf p}_0)$, the solution  in the collisionless case, is simply the following:
\begin{eqnarray}
f_i(\vec{\bf x},\vec{\bf p},t) = f_{i \, 0} {\bigg (}\vec{\bf x}_0(\vec{\bf x},\vec{\bf p},t), \vec{\bf p}_0(\vec{\bf x},\vec{\bf p},t)  {\bigg )}
\end{eqnarray}
The  current $\vec{\bf j}_i$ can  then be computed according to (\ref{eq_ckt_j}).

We next consider the case with a collision term of the form $C[f] = -\alpha\,  f_i + \beta $ where $\alpha(\vec{\bf x},\vec{\bf p},t), \beta(\vec{\bf x},\vec{\bf p},t)$ are both certain functions of position, momentum and time~\cite{Yan:2006ve}.  With this collision term, the formal exact solution to the kinetic equation is given by~\cite{Yan:2006ve}:
\begin{eqnarray} \label{eq_formal}
&&f_i(\vec{\bf x},\vec{\bf p},t)=f_{i0}(\vec{\bf x}_0,\vec{\bf p}_0)e^{-\int^{t}_{t_0}\alpha(\vec{\bf x}_{t'},\vec{\bf p}_{t'},{t'}) d{t'}}\nonumber\\
&&~~~~~~~~+\int^{t}_{t_0}\beta(\vec{\bf x}_s,\vec{\bf p}_s,s)e^{-\int^{t}_{s}\alpha(\vec{\bf x}_{t'},\vec{\bf p}_{t'},{t'}) d{t'}}ds. 
\end{eqnarray}
where again the $f_{i \, 0}$ is the initial condition. The $\vec{\bf x}_{\xi}(\vec{\bf x},\vec{\bf p},t)=\vec{\bf x}(\vec{\bf x},\vec{\bf p},t; \xi)=(x_{\xi}, y_{\xi}, z_{\xi}),~\vec{\bf p}_{\xi}(\vec{\bf x},\vec{\bf p},t)=\vec{\bf p}(\vec{\bf x},\vec{\bf p},t; \xi)=(p_{x_\xi}, p_{y_\xi}, p_{z_\xi})$ are the position and momentum at any intermediate time moment $\xi$ as determined by Eq.(\ref{eq:001}) and Eq.(\ref{eq:002}). Again the corresponding current can be computed according to (\ref{eq_ckt_j}). 
The familiar relaxation time approximation (RTA) is a special case of this form, with $\alpha \to 1/\tau_{R}$ (where $\tau_{R}$ is the relaxation time parameter) and $\beta \to f_{eq}/\tau_{R}$. It shall be emphasized that the local equilibrium distribution $f_{eq}=\frac{1}{e^{[\epsilon(\vec{\bf{p}})-\mu^*]/T^*}+1}$ here needs to be determined self-consistently at any spacetime point during the evolution, by fixing the local equilibrium parameters $T^*(\vec{\bf x},t),\mu^*(\vec{\bf x},t)$ from energy density and number  density. As such, the formal solution for the RTA case is an implicit one and needs to be numerically evaluated. 
The resulting current  in the RTA case  takes the following form:
\begin{eqnarray} \label{eq_RTA}
\vec{\bf j}_i^{\, RTA} 
&=&\int_{\vec{p}}\left [{\vec{\bf v}_p}+ q_i\, \vec{\bf B} \, \left(  {{\vec{\bf v}_p}}\cdot \vec{\bf b} \right) \right]
{\bigg [ } f_{i0}(\vec{\bf x}_0,\vec{\bf p}_0)e^{-\int^{t}_{t0}\frac{1}{\tau_{R}}d{t'}}\nonumber\\
&&~~~~ +\int^{t}_{t_0}\frac{1}{\tau_{R}}f_{eq}(\vec{\bf x}_s,\vec{p}_s,s)e^{-\int^{t}_{s}\frac{1}{\tau_{R}}d{t'}}ds {\bigg ]}  .  
\end{eqnarray}
In the above the current evolves from dominance of the ``memory'' of initial condition toward dominance of thermal equilibrium,  with $\tau_{R}$ controlling the time scale of such transition. It shall be noted that in a quark-gluon system created in a heavy ion collision there would be more than one relaxation time scales that are relevant to the kinetic, chemical, or spin equilibration processes. For the results to be presented in this paper we will assume just a single relaxation time scale for simplicity.

\section{Modeling Out-of-Equilibrium CME}

With the above obtained solutions to the chiral kinetic equation,  we now apply them for estimating the chiral magnetic current that can be generated during the early moments in heavy ion collisions when the created dense partonic matter is still out-of-equilibrium while the magnetic field is the strongest. In passing, we note that there has been study of pre-thermal chiral magnetic effect using classical-statistical field simulations~\cite{Mace:2016svc,Mace:2016shq,Fukushima:2015tza}. We also note that there has been attempt of applying chiral kinetic transport for describing long time evolution of the fireball assuming very long $\vec B$ field duration~\cite{Sun:2016nig}.

The partonic system at early time is characterized by the so-called saturation scale $Q_s$, on the order of $1\sim 3\rm GeV$ for RHIC and the LHC~\cite{Kowalski:2007rw}. We take $Q_s\simeq2\rm GeV$. According to recent kinetic studies of pre-equilibrium evolution (see e.g. \cite{Blaizot:2011xf,Blaizot:2013lga,Blaizot:2014jna}), while the system is initially gluon-dominated, the quarks are generated quickly on a time scale $\tau\sim 1/Q_s$ and then evolve toward thermal equilibrium.  We will use a formation time $\tau_{in}=0.1{\rm fm/c} \sim 1/Q_s$ as the starting time of our kinetic evolution of quark distributions until an end time of $\tau_f=0.6 {\rm fm/c}\sim 6/Q_s$ which is on the order of onset time for hydrodynamic evolution in heavy ion collisions. We use the following quark initial distributions at $\tau=\tau_{in}$:
\begin{eqnarray}
f_{i\, 0} =  \lambda_i  f_0 \,\, , \,\, f_0 = n_{ 0 } \, {\cal F}\left(|\vec{\bf p}|\right) \,  \exp\left[ -\frac{x^2}{R_x^2}   -\frac{y^2}{R_y^2}  -\frac{z^2}{R_z^2} \right]  \,\,
\end{eqnarray}
The spatial distribution is  Gaussian, with three width parameters. The longitudinal width is set as $R_z=1/(2Q_s)$.  The transverse widths are determined by nuclear geometry, e.g. for AuAu collisions (with nuclear radius $R_A\simeq 6.3\rm fm$) at impact parameter $b$, $R_x\to (R_A-b/2)$ and $R_y \to \sqrt{R_A^2-(b/2)^2}$. The initial momentum distribution ${\cal F}$ is not precisely known, so we will test the following three different  forms and compare the results:\\
(1) a Fermi-Dirac like form (FD) \\
${\cal F}_{FD} = \frac{1}{e^{(p-Q_s)/\Delta}+1}$ with $\Delta=0.2\rm GeV$; \\
(2) a soft-dominated Gaussian form (SG)  \\
${\cal F}_{SG} = e^{-\frac{p^2}{Q_s^2}}$; \\
(3) a hard-dominated Gaussian form (HG)  \\
${\cal F}_{HG} = (\frac{c\, p}{Q_s})^2\, e^{-(\frac{c\, p}{Q_s})^2}$ with $c=1.65$.   \\
The overall magnitude parameter $n_0$ is fixed by normalizing the quark number density at the fireball center via $\int_{\vec{\bf p}} f_0(\vec{\bf p},x=y=z=0) \to \xi Q_s^3$ (see e.g. \cite{Blaizot:2014jna})  and we will vary the parameter $\xi$ in a reasonable range to be consistent with that of typical initial condition used for hydrodynamic simulations.  Finally the constant $\lambda_i$ is used to specify the density difference across various species, namely  $u$, $\bar{u}$, $d$, and $\bar{d}$ quarks with positive/negative helicity respectively.
 We use the following choices: $\lambda_{u,+}=\lambda_{\bar{u},+}=\lambda_{d,+}=\lambda_{\bar{d},+}=1+\lambda_5$ while $\lambda_{u,-}=\lambda_{\bar{u},-}=\lambda_{d,-}=\lambda_{\bar{d},-}=1- \lambda_5$, where  $\lambda_5$ controls the initial imbalance of opposite helicity fermions. We will vary $\lambda_5$ to examine the dependence of pre-thermal CME effect on such initial imbalance. In a realistic heavy ion collision, the axial charge imbalance would be spatially fluctuating. Here the simple uniform imbalance is used to get a reasonable idea of how large the pre-thermal charge separation could be. With the presence of such imbalance, the relaxation time scale could become slightly different for fermions with opposite chirality. In this work we will use the same relaxation time scale for simplicity. Note also that the electric charge should be $q_u=-q_{\bar{u}}=\frac{2e}{3}$ and $q_d=-q_{\bar{d}}=-\frac{e}{3}$. For the rest of the paper we focus on the case of impact parameter $b=7.5 \rm fm$ corresponding roughly to $20-30\%$ centrality class.

\begin{figure}[!hbt]
\begin{center}
\includegraphics[scale=0.36]{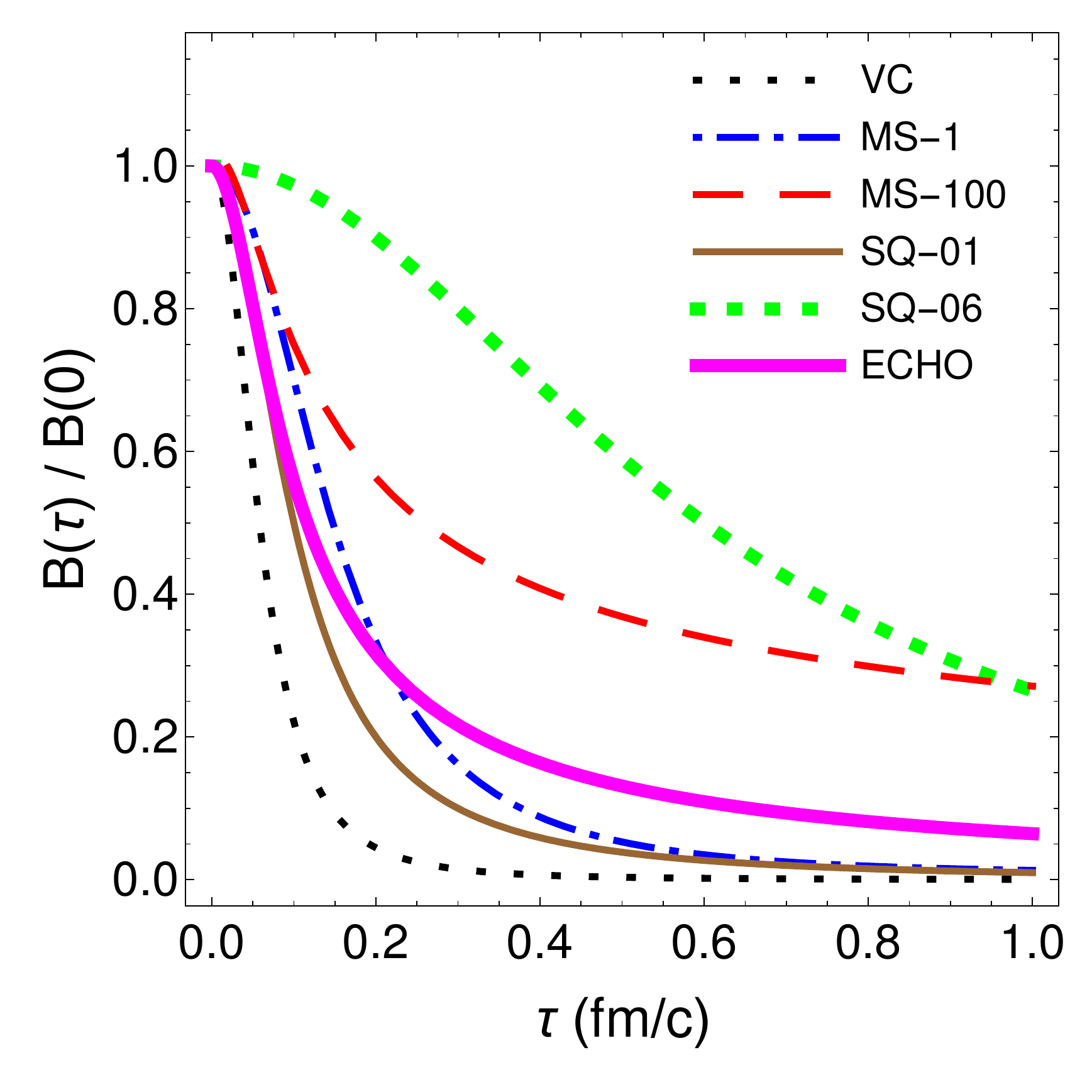}
\caption{Comparison of various time-dependent magnetic field $B(\tau)$ normalized by $B_0=B(\tau=0)$.}
\vspace{-0.5cm}
\label{fig_bt}
\end{center}
\end{figure}

The time evolution of magnetic field $\vec{\bf B}=B(\tau)\hat{\bf y}$ is not precisely determined. We will take an open attitude and compare a variety of possibilities proposed in the literature to provide a clear idea of the associated uncertainty. These include: \\
(a) The vacuum case  $B(\tau)=\frac{B_0}{\left[1+\left(\tau/\bar{\tau}\right)^2\right]^{3/2}}$ with $\bar{\tau} \simeq 0.076 \rm fm/c$ (referred to as VC hereafter)~\cite{Deng:2012pc,McLerran:2013hla}; \\
(b) The medium-modified case assuming a conductivity value equal to lattice computed thermal value at $T\simeq 1.45T_c$ (referred to as MS-1 hereafter)~\cite{McLerran:2013hla}; \\
(c) The medium-modified case assuming a conductivity value equal to 100 times the above mentioned lattice value (referred to as MS-100 hereafter)~\cite{McLerran:2013hla}; \\
(d) A widely used inverse-time-square parameterization $B(\tau)=\frac{B_0}{1+\left(\tau/\tau_B\right)^2}$ with $\tau_B = 0.1 \rm fm/c$ or $\tau_B=0.6 \rm fm/c$ (referred to as SQ-01 and SQ-06 hereafter)~\cite{Jiang:2016wve,Yin:2015fca,Yee:2013cya}; \\
(e) The dynamically evolving magnetic field $B(\tau)$ from the most recent magnetohydrodynamic computation based on ECHO-QGP code (referred to as ECHO hereafter)~\cite{Inghirami:2016iru}. \\
The peak value of $eB_0=6m_\pi^2$ at $\tau=0$ is set to be the same for all the above cases, taken from \cite{Bloczynski:2012en}. For a clear comparison, we show various $B(\tau)$ in Fig.~\ref{fig_bt}. Note that the magnetic field in general is not homogenous in space. Nevertheless from past event-by-event simulations (see e.g. \cite{Bloczynski:2012en})  such inhomogeneity of B field in most part of the overlapping zone in a heavy ion collision is quite mild (except near the edge). We will assume a uniform magnetic field which shall a reasonable approximation for gaining an idea of the magnitude for  the pre-thermal CME effect.  

\begin{figure}[!hbt]
\begin{center}
\includegraphics[scale=0.5]{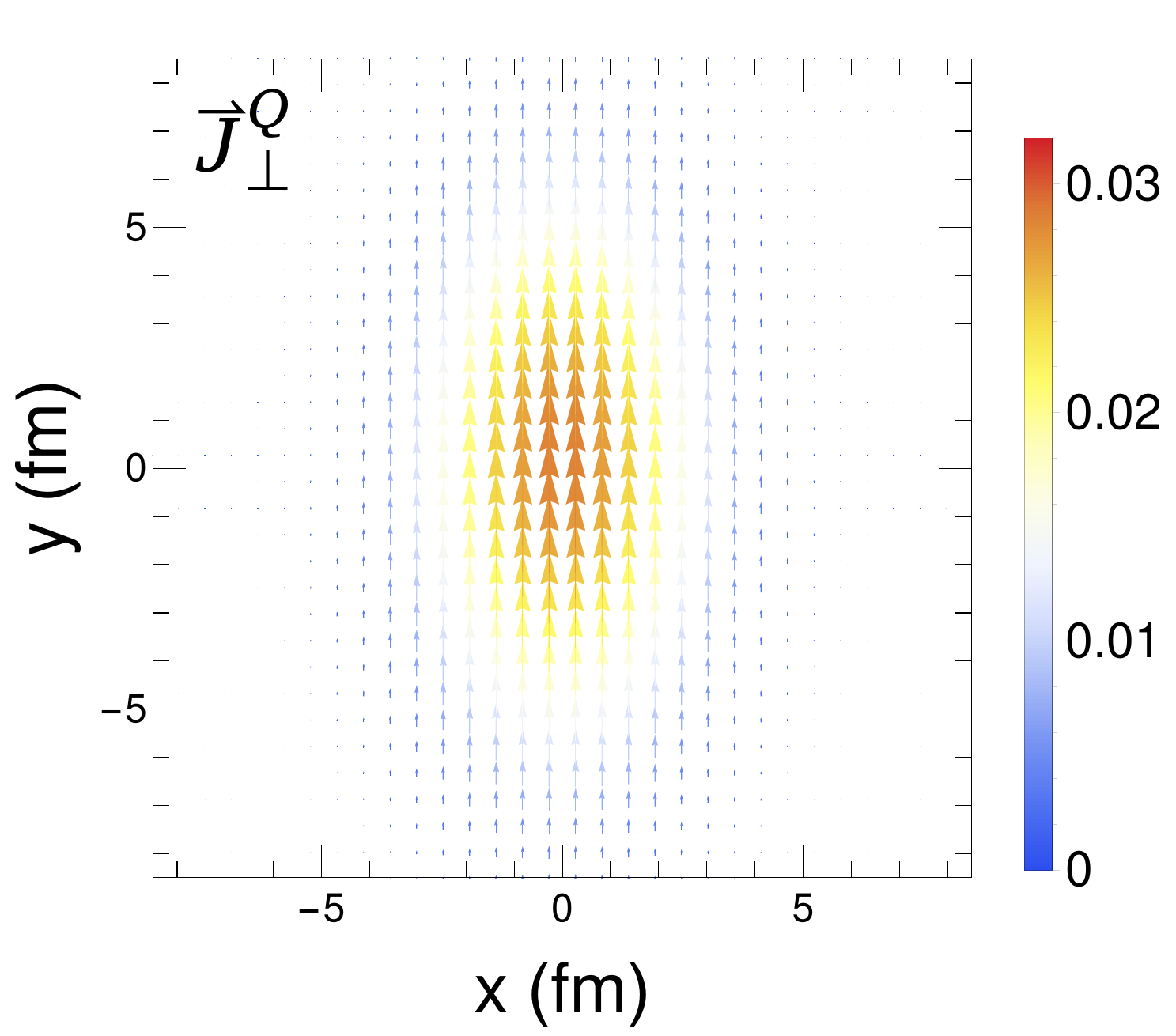}
\caption{Transverse charge current $\vec{J}^{\,Q}_\perp$ on the $x-y$ plane at time $\tau = 0.2 \rm fm/c$ (computed with FD initial distribution, ECHO magnetic field,  $\tau_R=0.1\rm fm/c$ and $\lambda_5=0.2$).}
\vspace{-0.5cm}
\label{fig_jy}
\end{center}
\end{figure}

\begin{figure}[!hbt]
\begin{center}
\includegraphics[scale=0.3]{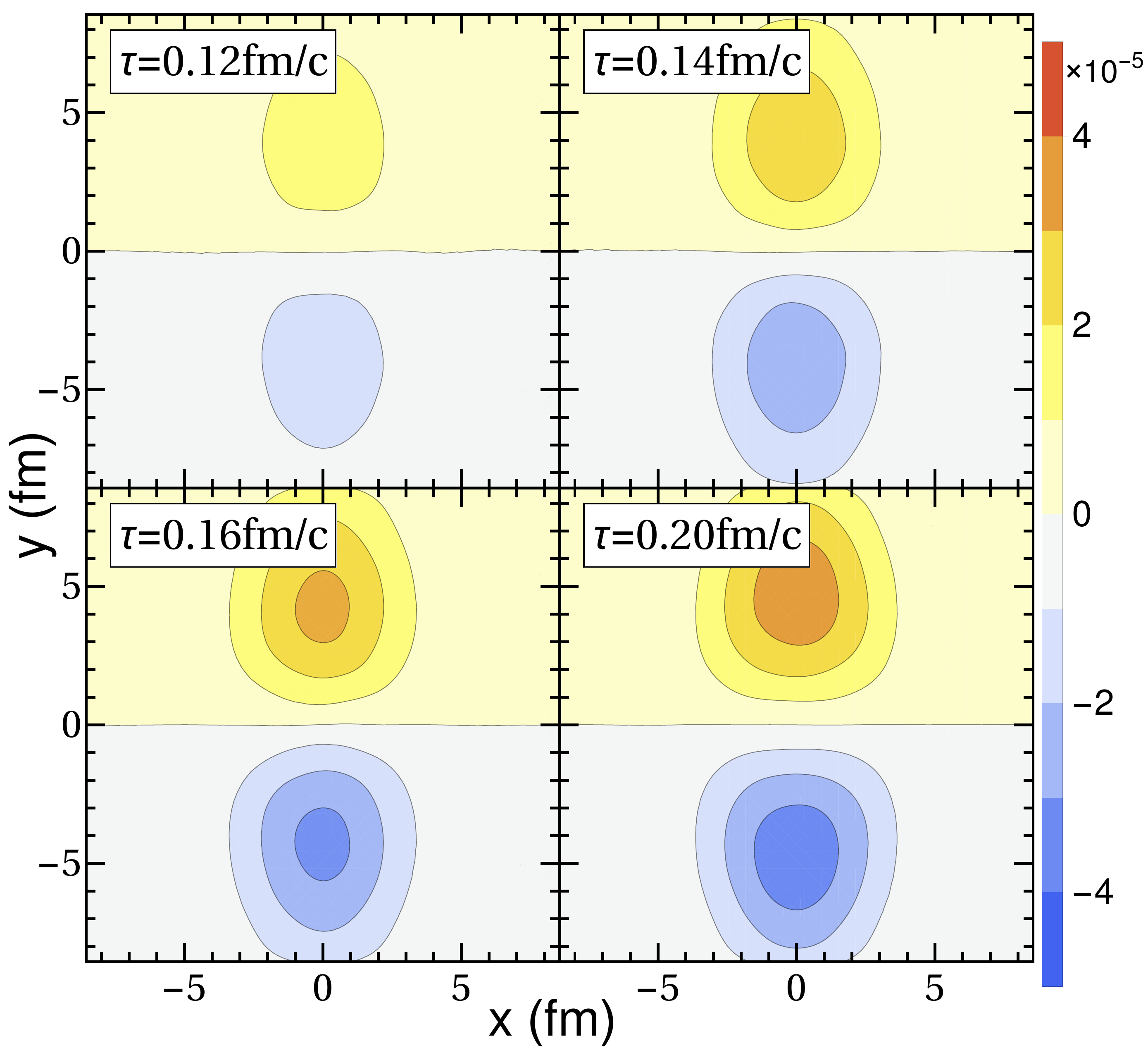}
\caption{Net charge density $n^{\,Q}$ (normalized by $\xi Q_s^3$) on the $x-y$ plane at different time  (computed with FD initial distribution, ECHO magnetic field, $\tau_R=0.1\rm fm/c$ and $\lambda_5=0.2$).}
\vspace{-0.5cm}
\label{fig_nq}
\end{center}
\end{figure}

Let us now demonstrate the out-of-equilibrium charge separation due to CME by examining the  transverse component $\vec{J}^{\,Q}_\perp$ of the electric charge current and the net charge density  distribution $n^Q$ on the $x-y$ plane from the chiral kinetic transport solutions. As already mentioned, the formal solution given in Eq.(\ref{eq_formal}) is implicit. To obtain concrete results, we've employed finite-difference numerical methods to explicitly evolve the kinetic equation in time. We use a spatial volume of 
$20\times 20\times 20\rm fm^3$ with grid size $\Delta r=\rm fm$ and a finite time step $\Delta t= 0.001 \rm fm/c$.   
In Fig.~\ref{fig_jy} we show the $\vec{J}^{\,Q}_\perp$ with the arrow indicating the direction of the current: it is evident that the current is aligned with magnetic field (along $y$-axis) and the magnitude is bigger in the area with larger local quark density. This CME-induced current will transport positive/negative charges in opposite direction and thus accumulate with time the separation of charges above/below the reaction plane. In Fig.~\ref{fig_nq} we show the net charge density distribution on the transverse plan at several time moments: indeed one clearly sees the gradual buildup of excessive positive charges on one side of the plane while negative charges on the other side, implying a growing charge separation with time.

\section{Results and Discussions}

\begin{figure}[!hbt]
\begin{center}
\includegraphics[scale=0.35]{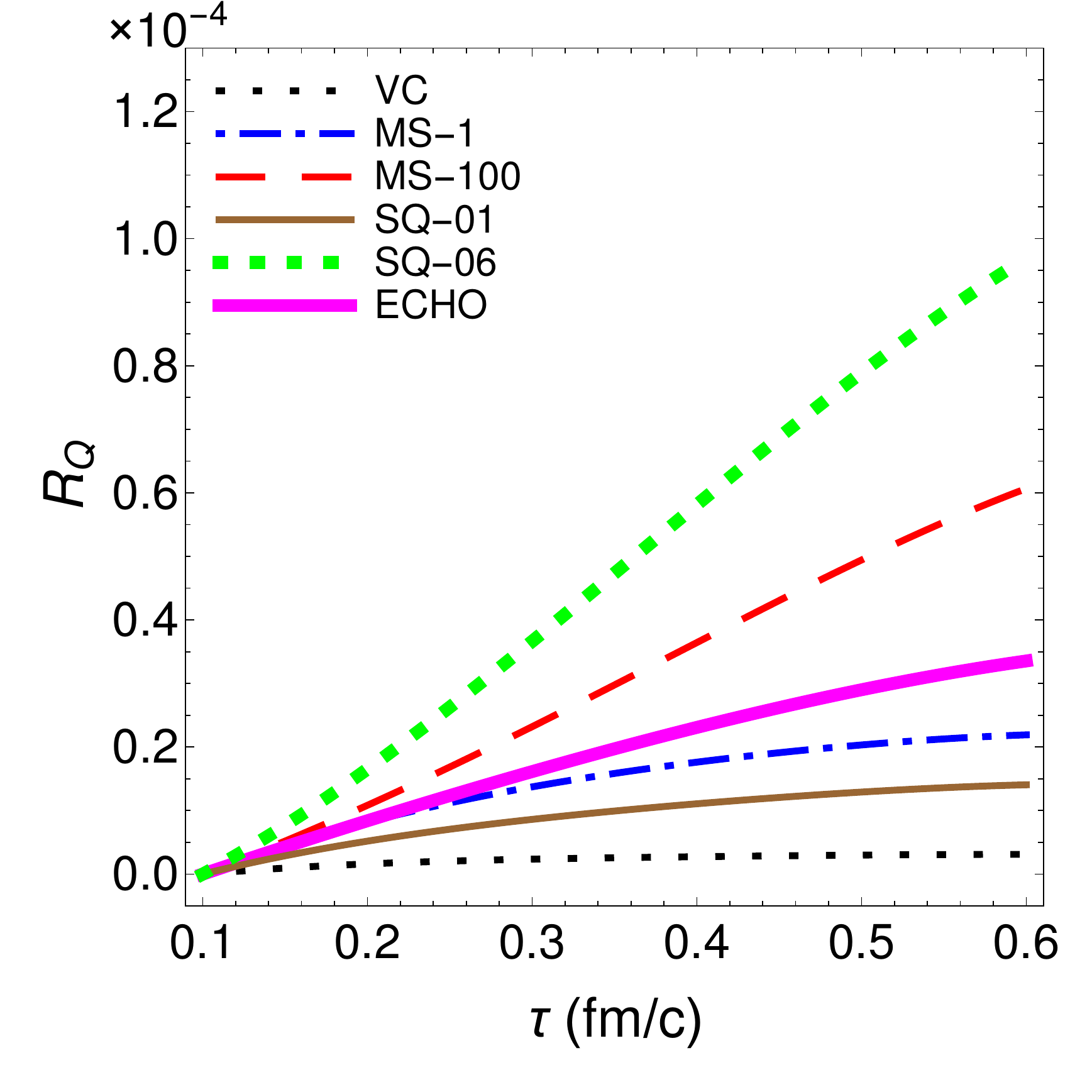}
\caption{The ratio $R_Q$ for quantifying charge separation across the reaction plane as a function of time, computed and compared for various choices of time-dependent magnetic field (with FD initial distribution, $\tau_R=0.1\rm fm/c$ and $\lambda_5=0.2$). }
\vspace{-0.5cm}
\label{fig_RQ}
\end{center}
\end{figure}

Let us then quantify the out-of-equilibrium charge separation effect and study its dependence on various ingredients in the modeling. To quantify this effect, we introduce a quantity $R_Q=N_Q/N_{total}$ defined as a ratio of the total number of net charge above reaction plane $N_Q$ (with equal number but opposite net charge $-N_Q$ below reaction plane) to the total number of quarks and anti-quarks $N_{total}$ in the system. This ratio $R_Q$ is shown in Fig.~\ref{fig_RQ} as a function of time $\tau$. The $R_Q$ monotonically grows with time in all cases, reflecting the accumulation of charge separation from continuous CME transport. One also finds a strong dependence of this effect on the magnetic field evolution with time. For example, the ECHO magnetic field (from magnetohydrodynamic simulations) could produce a charge separation  about {\em an order of magnitude larger} than the vacuum case. It is therefore crucial to treat magnetic field as dynamically evolving by  properly accounting for medium feedback.

\begin{figure}[!hbt]
\begin{center}
\includegraphics[scale=0.35]{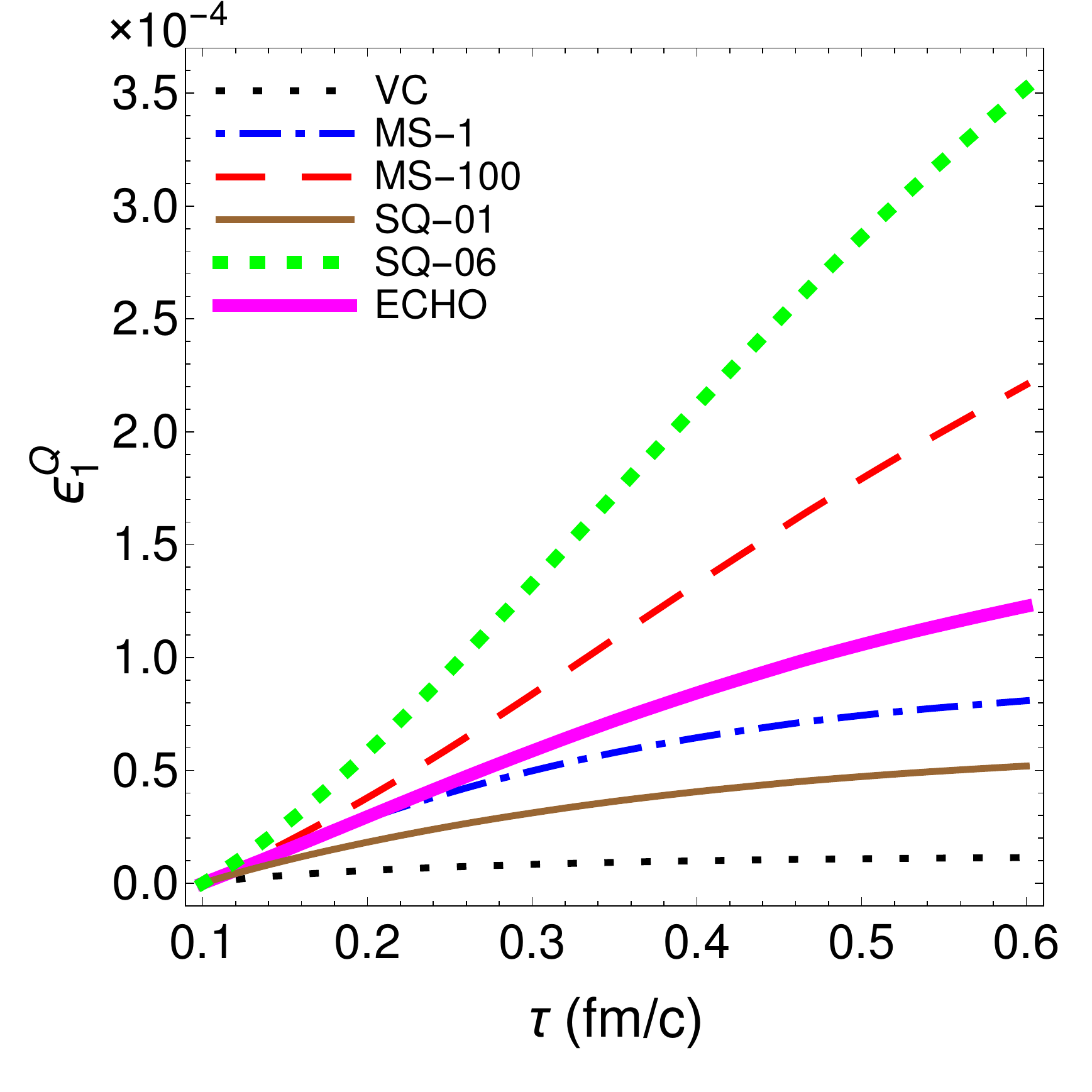}
\caption{The charge dipole moment $\epsilon_1^Q$ for quantifying charge separation as a function of time, computed and compared for various choices of time-dependent magnetic field (with FD initial distribution, $\tau_R=0.1\rm fm/c$ and $\lambda_5=0.2$).}
\vspace{-0.5cm}
\label{fig_nq_FD}
\end{center}
\end{figure}

Another way to quantify this CME-induced pre-thermal charge separation, is to define a weighed charge dipole moment of the net charge density distribution  on the transverse plane, as follows:
\begin{eqnarray}
\epsilon_1^Q = \frac{\int dz dr_\perp^2 d\phi  \, r_\perp^2 \, \sin{\phi} \, n^Q}{\int dz dr_\perp^2 d\phi  \, r_\perp^2 \, n^{tot.}}
\end{eqnarray}
where $r_\perp$ and $\phi$ are transverse radial and azimuthal coordinates on $x-y$ plane, $n^Q$ is the net charge density while $n^{tot.}$ is the total quark and anti-quark number density providing the normalization in defining the dimensionless dipole moment $\epsilon_1^Q$.  In Fig.~\ref{fig_nq_FD} we show $\epsilon_1^Q$ as a function of time. In consistency with $R_Q$ shown in Fig.~\ref{fig_RQ}, the dipole moment is found to grow with time and sensitively depends on the time evolution of the magnetic field.

\begin{figure}[!hbt]
\begin{center}
\includegraphics[scale=0.35]{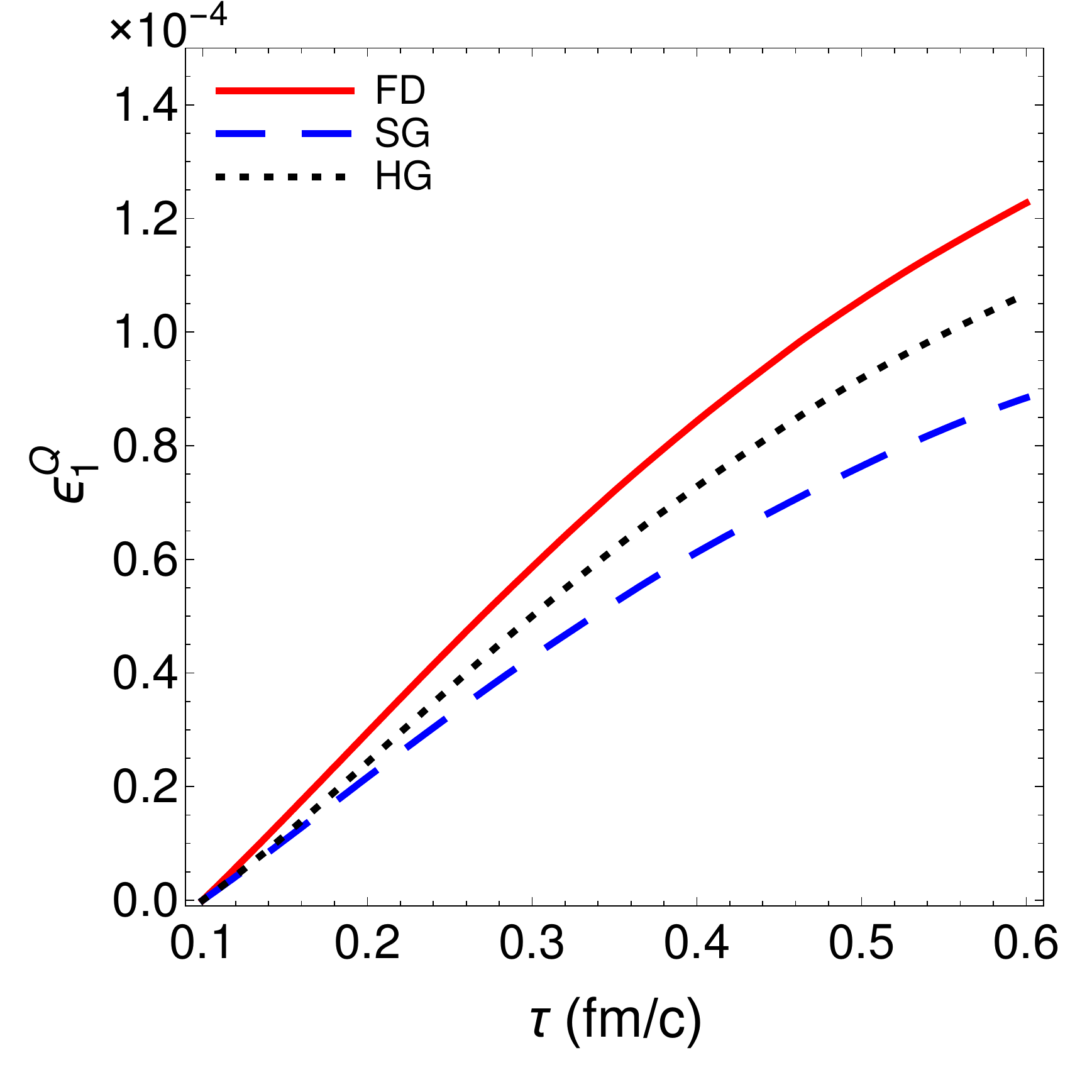}
\caption{The charge dipole moment $\epsilon_1^Q$  as a function of time, computed and compared for three choices (FD, SG, HG) of initial momentum distribution  (with ECHO magnetic field, $\tau_R=0.1\rm fm/c$ and $\lambda_5=0.2$).}
\vspace{-0.5cm}
\label{fig_compare_f}
\end{center}
\end{figure}

We next examine the influence of the initial momentum distribution on the resulting charge separation dipole. To do that, we compare $\epsilon_1^Q$ as a function of $\tau$ computed from three different initial distributions (FD, SG, HG): see Fig.~\ref{fig_compare_f}. Note that for the three distributions the total quark/antiquark number is normalized to be the same for fair comparison. The results demonstrate a mild dependence on such initial distribution, and appear to indicate that the charge separation could be enhanced if the momentum is more distributed in the soft regime.

\begin{figure}[!hbt]
\begin{center}
\includegraphics[scale=0.35]{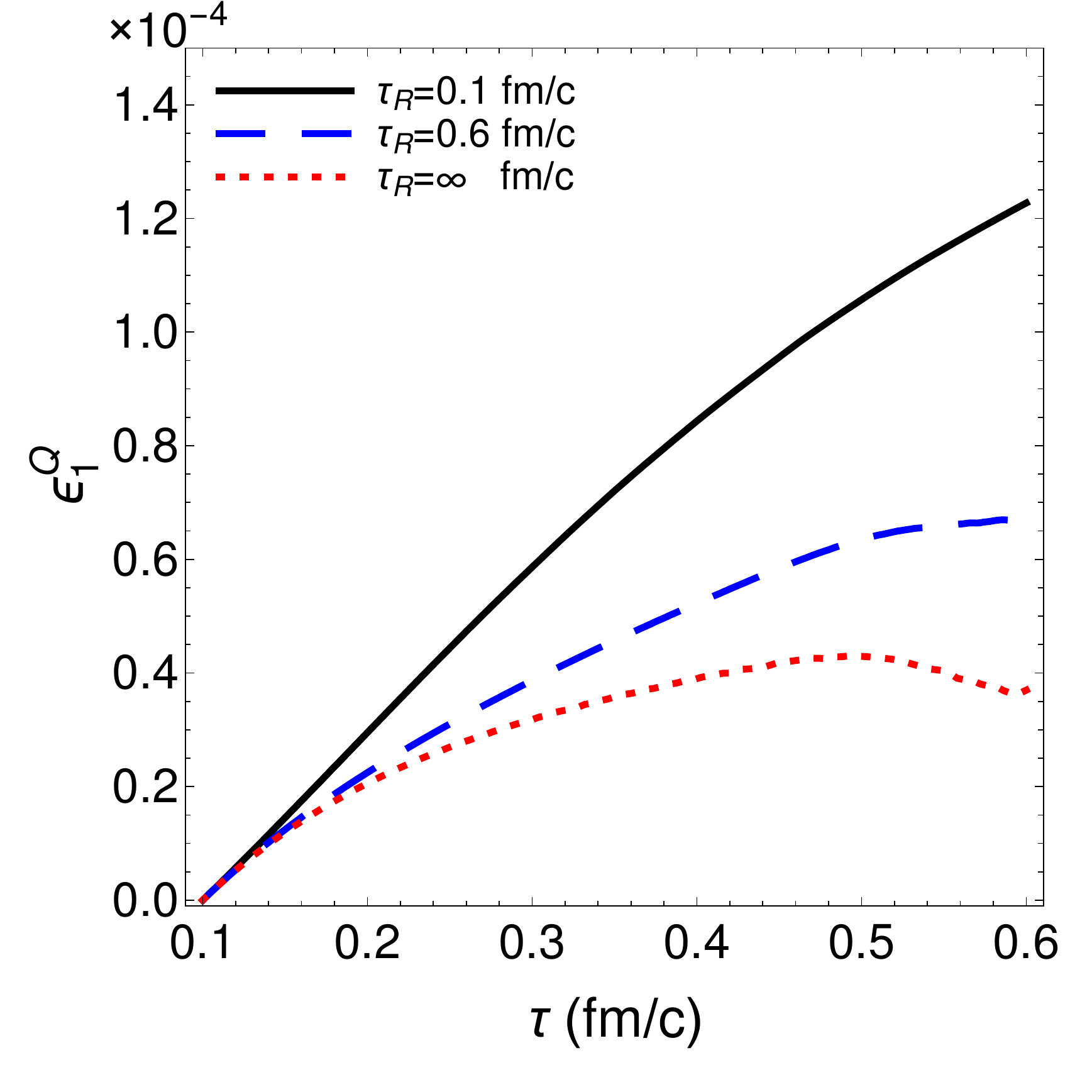}
\caption{The charge dipole moment $\epsilon_1^Q$  as a function of time, computed and compared for three choices  of relaxation time $\tau_R$  (with ECHO magnetic field, FD initial distribution and  $\lambda_5=0.2$).}
\vspace{-0.5cm}
\label{fig_compare_relax}
\end{center}
\end{figure}

Finally we study the influence of the relaxation parameter $\tau_R$ on the resulting charge separation dipole. In Fig.~\ref{fig_compare_relax} the $\epsilon_1^Q$ as a function of $\tau$ is computed for three choices of $\tau_R$: collision-less limit $\tau_R=\infty$, slow relaxation case $\tau_R=0.6\rm fm/c$ and fast relaxation case $\tau_R=0.1\rm fm/c$. The comparison clearly demonstrates that the charge separation increases with decreasing relaxation time (i.e. stronger scattering in the system). This may be understood as follows: more scattering would prevent the quarks/antiquarks from streaming away too quickly thus keeping their density higher  to generate more contributions to the anomalous current. Such pre-thermal CME current, however, originates from quantum response of chiral fermions to the external magnetic field and the relaxation affects it only in an indirect way, which is different from the situation of the usual pre-thermal collective flow where interesting universal scaling behavior occurs~\cite{Vredevoogd:2008id,Keegan:2016cpi}.

So far we have not discussed the influence of  two  other important ingredients:   the initial helicity imbalance parameterized by $\lambda_5$ and the peak magnetic field strength $B_0$. In fact the dependence is fairly simple: the charge separation (described by either $R_Q$ or $\epsilon^1_Q$) is found to be simply linearly proportional to both of these factors. This conclusion should though be put in a context, as the homologous initial chirality imbalance $\lambda_5$ used in this work is only an approximate implementation of axial charge dynamics. A dynamical treatment of axial charge would involve inhomogeneous generation from gluon topological fluctuations as well as possible chiral plasma instability~\cite{Akamatsu:2013pjd}.

It may be noted that the obtained pre-thermal CME-induced charge separation effect has been found to be rather small for our current choice of parameters. This may be due to a number of factors. Firstly the total evolution time (from $0.1\rm fm/c$ to $0.6 \rm fm/c$) is fairly short, which limits the accumulation of charge separation. Secondly, in the present formulation of chiral kinetic theory the anomalous term responds instantaneously to the applied magnetic field, so the resulting anomalous current decreases in time rapidly along with the magnetic field thus hindering the buildup of charge separation. This is an important and challenging issue that requires further   investigation. Finally, the imbalance in number density between opposite helicity fermions (that we currently use in these calculations) is small compared with the relevant system scale $Q_s$ which is analogous to the situation of a very small ratio for axial charge density $n_5$ to entropy density in the thermal case.

We end by discussing the implication for the subsequent hydrodynamic evolution stage. The occurrence of the pre-hydro CME implies that by the time of hydro onset, the fermion density and current can  no longer be set as trivially vanishing (as usually done). Instead, as shown in Figs.~\ref{fig_jy} and \ref{fig_nq}, the per-thermal transport induces a nontrivial initial condition by that time and should  be incorporated into the subsequent hydrodynamic evolution  as nontrivial initial conditions for the various fermion currents including both the zeroth component (i.e. density) and the spatial component (vector 3-current density). With the recently developed Anomlaous-Viscous Fluid Dynamics (AVFD) framework~\cite{Jiang:2016wve}, we have tested and demonstrated that indeed such pre-hydro charge separation can be built into hydro initial conditions and will survive through the hydro stage to contribute to the final hadron observables.

\section{Summary}

In summary, we have performed a first phenomenological study of the CME-induced charge separation during the pre-thermal stage in heavy ion collisions. Such study will be very important for addressing the  crucial challenge due to the fact that the extremely strong magnetic field in such collisions may last only for a brief moment and the CME current may have to occur at so early a stage that the quark-gluon matter is still far from thermal equilibrium. Utilizing the tool of chiral kinetic theory, we have developed analytic solutions for the collision-less limit and the relaxation time approximation. We have quantified the net charge dipole moment arising from the per-thermal CME-induced charge separation and studied its dependence on various ingredients in the modeling. The effect is found to be very sensitive to the time dependence of the magnetic field and also influenced by the initial quark momentum spectrum as well as the relaxation time of the system evolution toward thermal equilibrium. Within the present approach, such pre-thermal charge separation is found to be modest. Finally the implication of pre-thermal CME for the subsequent hydrodynamic evolution has been discussed.

With this proof-of-concept study, we aim to develop in the future a more comprehensive and realistic pre-thermal CME modeling tool that will also be seamlessly integrated with an anomalous hydrodynamic evolution. One lesson we've learned from the present study is that the magnetic field driven effects may be significant during the early stage when the field is most strong. For example the pre-thermal CME leads to nontrivially modified initial conditions for hydrodynamic evolution. In addition to such transport effects, it could be anticipated that the early time magnetic field may have an even stronger and more direct influence on the particles that are dominantly produced through initial hard processes~\cite{Guo:2015nsa,Basar:2014swa}, such as the high momentum quarkonia as well as  dileptons and photons. During the formation of these particles at the earliest moments after a collision, the strong magnetic field (and the anomalous transport of partons driven by it) may possibly leave an imprint in their production, which would be an interesting  topic for future study.


\vspace{0.2in}

 {\bf Acknowledgments.}
The authors thank K.~Fukushima, X.~G.~Huang, M.~Stephanov and H.~U.~Yee for helpful discussions. The research of AH and PZ is supported by the NSFC and MOST Grant Nos. 11335005, 11575093, 2013CB922000 and 2014CB845400.  YJ is supported by the DFG Collaborative Research Center  ``SFB 1225 (ISOQUANT)''. This material is partly based upon work supported by the U.S. Department of Energy, Office of Science, Office of Nuclear Physics, within the framework of the Beam Energy Scan Theory (BEST) Topical Collaboration. JL and SS are also supported in part by the National Science Foundation under Grant No. PHY-1352368.


\begin{thebibliography}{99}

\bibitem{Adler:1969gk}
  S.~L.~Adler,
  Phys.\ Rev.\  {\bf 177}, 2426 (1969).

\bibitem{Bell:1969ts}
  J.~S.~Bell and R.~Jackiw,
  Nuovo Cim.\ A {\bf 60}, 47 (1969).


\bibitem{Kharzeev:2004ey}
  D.~Kharzeev,
  Phys.\ Lett.\ B {\bf 633}, 260 (2006).

\bibitem{Kharzeev:2007tn}
  D.~Kharzeev and A.~Zhitnitsky,
  Nucl.\ Phys.\  A {\bf 797}, 67 (2007).

 \bibitem{Kharzeev:2007jp}
  D.~E.~Kharzeev, L.~D.~McLerran and H.~J.~Warringa,
  Nucl.\ Phys.\  A {\bf 803}, 227 (2008).

\bibitem{Fukushima:2008xe}
  K.~Fukushima, D.~E.~Kharzeev and H.~J.~Warringa,
  Phys.\ Rev.\ D {\bf 78}, 074033 (2008).


\bibitem{Kharzeev:2010gd}
  D.~E.~Kharzeev and H.~U.~Yee,
  Phys.\ Rev.\ D {\bf 83}, 085007 (2011)
  doi:10.1103/PhysRevD.83.085007
  [arXiv:1012.6026 [hep-th]].

  \bibitem{Burnier:2011bf}
  Y.~Burnier, D.~E.~Kharzeev, J.~Liao and H.~U.~Yee,
  Phys.\ Rev.\ Lett.\  {\bf 107}, 052303 (2011); arXiv:1208.2537 [hep-ph].

  %
\bibitem{Son:2009tf}
  D.~T.~Son and P.~Surowka,
  Phys.\ Rev.\ Lett.\  {\bf 103}, 191601 (2009).


\bibitem{Kharzeev:2010gr}
  D.~E.~Kharzeev and D.~T.~Son,
  Phys.\ Rev.\ Lett.\  {\bf 106}, 062301 (2011).
  
\bibitem{Sadofyev:2010is} 
  A.~V.~Sadofyev, V.~I.~Shevchenko and V.~I.~Zakharov,
  Phys.\ Rev.\ D {\bf 83}, 105025 (2011). 

\bibitem{Landsteiner:2011iq} 
  K.~Landsteiner, E.~Megias, L.~Melgar and F.~Pena-Benitez,
  JHEP {\bf 1109}, 121 (2011). 
  
\bibitem{Jiang:2015cva}
  Y.~Jiang, X.~G.~Huang and J.~Liao,
  Phys.\ Rev.\ D {\bf 92}, no. 7, 071501 (2015).

\bibitem{STAR_LPV1} B.~I.~Abelev {\it et al.} (STAR Collaboration), Phys. Rev. Lett. {\bf 103} (2009) 251601.

\bibitem{STAR_LPV_BES}
L. Adamczyk {\it et al.} (STAR Collaboration), Phys. Rev. Lett. {\bf 113} (2014) 052302.

\bibitem{ALICE_LPV}
B. I. Abelev {\it et al.} (ALICE Collaboration), Phys. Rev. Lett. {\bf 110} (2013) 021301.

 \bibitem{Adamczyk:2015eqo}
  L.~Adamczyk {\it et al.} [STAR Collaboration],
  Phys.\ Rev.\ Lett.\  {\bf 114}, no. 25, 252302 (2015).

\bibitem{Li:2014bha}
  Q.~Li {\it et al.},
  Nature Phys.\  {\bf 12}, 550 (2016)
  doi:10.1038/nphys3648
  [arXiv:1412.6543 [cond-mat.str-el]].

\bibitem{Xiong:2015nna}
  J.~Xiong, S.~K.~Kushwaha, T.~Liang, J.~W.~Krizan, W.~Wang, R.~J.~Cava and N.~P.~Ong,
  arXiv:1503.08179 [cond-mat.str-el].


  \bibitem{2015PhRvX...5c1023H} Huang, X., Zhao, L., Long, Y., et al.\ 2015, Physical Review X, 5, 031023.

  \bibitem{2016NatCo...711615A} Arnold, F., Shekhar, C., Wu, S.-C., et al.\ 2016, Nature Communications, 7, 11615.



\bibitem{Kharzeev:2015znc}
  D.~E.~Kharzeev, J.~Liao, S.~A.~Voloshin and G.~Wang,
  Prog.\ Part.\ Nucl.\ Phys.\  {\bf 88}, 1 (2016).

\bibitem{Kharzeev:2015kna}
  D.~E.~Kharzeev,
  Ann.\ Rev.\ Nucl.\ Part.\ Sci.\  {\bf 65}, 193 (2015)
  doi:10.1146/annurev-nucl-102313-025420
  [arXiv:1501.01336 [hep-ph]].

\bibitem{Liao:2014ava}
  J.~Liao,
  Pramana {\bf 84}, no. 5, 901 (2015)
  [arXiv:1401.2500 [hep-ph]].

\bibitem{burkov}
A.~A.~Burkov,
J. Phys.: Condens. Matter {\bf 27}, 113201  (2015) .


\bibitem{Bzdak:2011yy}
  A.~Bzdak and V.~Skokov,
  Phys.\ Lett.\ B {\bf 710}, 171 (2012).



\bibitem{Deng:2012pc}
  W.~T.~Deng and X.~G.~Huang,
  Phys.\ Rev.\ C {\bf 85}, 044907 (2012).

\bibitem{Bloczynski:2012en}
  J.~Bloczynski, X.~G.~Huang, X.~Zhang and J.~Liao,
  Phys.\ Lett.\ B {\bf 718}, 1529 (2013).


\bibitem{McLerran:2013hla}
  L.~McLerran and V.~Skokov,
  Nucl.\ Phys.\ A {\bf 929}, 184 (2014).

\bibitem{Gursoy:2014aka}
  U.~Gursoy, D.~Kharzeev and K.~Rajagopal,
  Phys.\ Rev.\ C {\bf 89}, no. 5, 054905 (2014).


\bibitem{Tuchin:2015oka}
  K.~Tuchin,
  Phys.\ Rev.\ C {\bf 93}, no. 1, 014905 (2016).

\bibitem{Li:2016tel}
  H.~Li, X.~l.~Sheng and Q.~Wang,
  Phys.\ Rev.\ C {\bf 94}, no. 4, 044903 (2016)

\bibitem{Inghirami:2016iru}
  G.~Inghirami, L.~Del Zanna, A.~Beraudo, M.~H.~Moghaddam, F.~Becattini and M.~Bleicher,
  Eur.\ Phys.\ J.\ C {\bf 76}, no. 12, 659 (2016).

\bibitem{Guo:2015nsa}
  X.~Guo, S.~Shi, N.~Xu, Z.~Xu and P.~Zhuang,
  Phys.\ Lett.\ B {\bf 751}, 215 (2015)

  \bibitem{Jiang:2016wve}
  Y.~Jiang, S.~Shi, Y.~Yin and J.~Liao,
  arXiv:1611.04586 [nucl-th].

\bibitem{Yin:2015fca}
  Y.~Yin and J.~Liao,
  Phys.\ Lett.\ B {\bf 756}, 42 (2016)

\bibitem{Hirono:2014oda}
  Y.~Hirono, T.~Hirano and D.~E.~Kharzeev,
  arXiv:1412.0311 [hep-ph].


\bibitem{Yee:2013cya}
  H.~U.~Yee and Y.~Yin,
  Phys.\ Rev.\ C {\bf 89}, no. 4, 044909 (2014).


\bibitem{Stephanov:2012ki}
  M.~A.~Stephanov and Y.~Yin,
  Phys.\ Rev.\ Lett.\  {\bf 109}, 162001 (2012).

\bibitem{Son:2012wh}
  D.~T.~Son and N.~Yamamoto,
  Phys.\ Rev.\ Lett.\  {\bf 109}, 181602 (2012).

\bibitem{Son:2012zy} 
  D.~T.~Son and N.~Yamamoto,
  Phys.\ Rev.\ D {\bf 87}, no. 8, 085016 (2013). 
  
\bibitem{Chen:2014cla} 
  J.~Y.~Chen, D.~T.~Son, M.~A.~Stephanov, H.~U.~Yee and Y.~Yin,
  Phys.\ Rev.\ Lett.\  {\bf 113}, no. 18, 182302 (2014). 
  
\bibitem{Kharzeev:2016sut} 
  D.~E.~Kharzeev, M.~A.~Stephanov and H.~U.~Yee,
  Phys.\ Rev.\ D {\bf 95}, no. 5, 051901 (2017)
  
\bibitem{Chen:2012ca}
  J.~W.~Chen, S.~Pu, Q.~Wang and X.~N.~Wang,
  Phys.\ Rev.\ Lett.\  {\bf 110}, no. 26, 262301 (2013).

\bibitem{Hidaka:2016yjf} 
  Y.~Hidaka, S.~Pu and D.~L.~Yang,
  arXiv:1612.04630 [hep-th].


  
\bibitem{Mueller:2017lzw}
  N.~Mueller and R.~Venugopalan,
  arXiv:1701.03331 [hep-ph];
  arXiv:1702.01233 [hep-ph].

\bibitem{Gorbar:2017cwv} 
  E.~V.~Gorbar, V.~A.~Miransky, I.~A.~Shovkovy and P.~O.~Sukhachov,
  arXiv:1702.02950 [cond-mat.mes-hall].
  
\bibitem{Yan:2006ve}
  L.~Yan, P.~Zhuang and N.~Xu,
  Phys.\ Rev.\ Lett.\  {\bf 97}, 232301 (2006).


\bibitem{Mace:2016svc}
  M.~Mace, S.~Schlichting and R.~Venugopalan,
  Phys.\ Rev.\ D {\bf 93}, no. 7, 074036 (2016)

  \bibitem{Mace:2016shq}
  M.~Mace, N.~Mueller, S.~Schlichting and S.~Sharma,
  Phys.\ Rev.\ D {\bf 95}, no. 3, 036023 (2017).

\bibitem{Fukushima:2015tza}
  K.~Fukushima,
  Phys.\ Rev.\ D {\bf 92}, no. 5, 054009 (2015).

\bibitem{Sun:2016nig}
  Y.~Sun, C.~M.~Ko and F.~Li,
  Phys.\ Rev.\ C {\bf 94}, no. 4, 045204 (2016).
  Y.~Sun and C.~M.~Ko,
  arXiv:1612.02408 [nucl-th].

\bibitem{Kowalski:2007rw}
  H.~Kowalski, T.~Lappi and R.~Venugopalan,
  Phys.\ Rev.\ Lett.\  {\bf 100}, 022303 (2008).



\bibitem{Blaizot:2011xf}
  J.~P.~Blaizot, F.~Gelis, J.~F.~Liao, L.~McLerran and R.~Venugopalan,
  Nucl.\ Phys.\ A {\bf 873}, 68 (2012).

\bibitem{Blaizot:2013lga}
  J.~P.~Blaizot, J.~Liao and L.~McLerran,
  Nucl.\ Phys.\ A {\bf 920}, 58 (2013).

\bibitem{Blaizot:2014jna}
  J.~P.~Blaizot, B.~Wu and L.~Yan,
  Nucl.\ Phys.\ A {\bf 930}, 139 (2014).


\bibitem{Vredevoogd:2008id} 
  J.~Vredevoogd and S.~Pratt,
  Phys.\ Rev.\ C {\bf 79}, 044915 (2009). 
  
\bibitem{Keegan:2016cpi} 
  L.~Keegan, A.~Kurkela, A.~Mazeliauskas and D.~Teaney,
  JHEP {\bf 1608}, 171 (2016). 
  
\bibitem{Akamatsu:2013pjd} 
  Y.~Akamatsu and N.~Yamamoto,
  Phys.\ Rev.\ Lett.\  {\bf 111}, 052002 (2013).

\bibitem{Basar:2014swa}
  G.~Basar, D.~E.~Kharzeev and E.~V.~Shuryak,
  Phys.\ Rev.\ C {\bf 90}, no. 1, 014905 (2014).


\end{thebibliography}
\end{document}